\documentclass{article}  
\usepackage{jamaica04}
\frompage{000} \topage{000}                                              

\def\sqrtsNN{\mbox{$\sqrt{s_\mathrm{NN}}$}}
\def\GeVc{\mbox{$\mathrm{GeV}/c$}}
\def\lt{\mbox{$<$}}

\newcommand{ \be }{\begin{equation}}    
\newcommand{ \ee }{\end{equation}}    
\newcommand{ \bea }{\begin{eqnarray}}    
\newcommand{ \eea }{\end{eqnarray}}    
\newcommand{ \la }{\langle}    
\newcommand{ \ra }{\rangle}

\usepackage{color}

\def\rightmark{Anisotropic flow at RHIC}
\def\leftmark{A. H. Tang for the STAR collaboration}
 
\title{Anisotropic flow at RHIC} 
\authors{
{A. H. Tang $^1$ for the STAR Collaboration 
}\\[2.812mm]
{\normalsize
\hspace*{-8pt}$^1$ NIKHEF and Brookhaven National Lab, \\ 
Physics Department, P.O. Box 5000, Upton, NY 11973, USA\\[0.2ex] 
}}

\abstract{
We present the first measurement of directed flow ($v_1$) at the 
Relativistic Heavy Ion Collider (RHIC). $v_1$ is found to be consistent 
with zero at pseudorapidities $\eta$ from 
$-1.2$ to $1.2$, then rises to the level of a couple of percent over the 
range  $2.4 < |\eta| < 4$.  The latter observation is similar to that from 
NA49 if the SPS rapidities are shifted by the difference in beam rapidity 
between RHIC and SPS. We studied the evolution of elliptic flow from p+p 
collisions through d+Au collision, and onto Au+Au collisions. 
Measurements of higher harmonics are presented and discussed. 
 }


\begin{document}
 
\maketitle
\setcounter{page}{1}

\vspace*{-0.5cm} 
\section{Introduction}\label{intro}
\vspace*{-0.1cm} 
Study of anisotropic flow is widely recognized as an important tool 
to probe the hot, dense matter that is created by the heavy ion 
collisions~\cite{Reisdorf}. Anisotropic flow means that, in non-central 
heavy ion collisions, the azimuthal distribution of outgoing particles with 
respect to the reaction plane is not uniform. It can be 
characterized~\cite{methodPaper} by Fourier coefficients
\be
v_{n} = \la  \cos n (\phi - \psi) \ra
\label{vndef}
\ee
where $\phi$ denotes the azimuthal angle of an outgoing particle, $\psi$ is
the orientation of the reaction plane, and $n$ denotes the harmonic. The
first Fourier coefficient, $v_1$, referred to as {\it directed flow},
describes the sideward motion of the fragments 
in ultra-relativistic nuclear collisions and it carries early 
information from the collision. Its shape at midrapidity is of special 
interest because it might reveal a signature of a possible phase transition 
from normal nuclear matter to a quark-gluon plasma~\cite{wiggle}.
Elliptic flow ($v_2$) is caused by the initial geometric deformation of 
the reaction region in the transverse plane. At low transverse momentum, 
roughly speaking, large values of elliptic flow are considered signatures
of hydrodynamic behavior. At large transverse momentum, in a {\it jet quenching} 
picture~\cite{Wang01}, elliptic flow results from jets emitted 
out-of-plane suffer more energy loss than those emitted in-plane. 
Higher harmonics reflect the details of the initial geometry. 
Recently it is reported~\cite{Kolb} that the magnitude and even the 
sign of $v_4$ are more sensitive than $v_2$ to initial conditions in 
the hydrodynamic calculations.

\vspace*{-0.1cm}  
\section{Data set}
\vspace*{-0.1cm} 
The data come from the second year of operation of Relativistic Heavy Ion 
collider (RHIC) at its top energy $\sqrtsNN = 200$ GeV. 
The STAR detector~\cite{STAR} main Time Projection Chamber 
(TPC~\cite{STARTPC}) and two forward TPCs (FTPC~\cite{STARFTPC}) were 
used in the analysis. The data set consists of about 2 million   
minimum bias and 1.2 million central trigger Au+Au events  
, 7 million d+Au minimum bias events and 11 million p+p minimum bias 
events. For $v_1$ analyses there were 70 thousand events available which 
included the FTPCs. The centrality definition in this paper is the same as 
used previously by STAR~\cite{Adler:2002xw}. Tracks used to reconstruct the 
flow vector, or generating function in case of cumulant method, were subject 
to the same quality cuts that were used in $\sqrtsNN = 130$ ~GeV 
analysis~\cite{Adler:2002pu}, except for the 
low transverse momentum cutoff, which for this analysis is 0.15 GeV/c instead 
of 0.1 GeV/c. For the scalar product analysis (introduced later in this 
paper), a tight cut on $\eta$ (from --1. to 1.) is applied on the flow 
vector, as well as a tight cut on distance of the closest approach (DCA) 
(from 0 to 1 cm).
\vspace*{-0.1cm} 
\section{Results}
\vspace*{-0.1cm}  
\subsection{Directed flow at RHIC}\label{details}
\vspace*{-0.1cm} 
The difficulties in studying directed flow are
that the signal is small and the non-flow contribution to the
two-particle azimuthal correlations can be comparable or even larger
than the correlations due to flow.  
We use the three-particle cumulant method~\cite{Borghini} and event plane 
method with mixed harmonics~\cite{methodPaper}  in $v_1$ analysis and the 
results agree with each other~\cite{MarkusPoster}. Both methods are less 
sensitive to two-particle non-flow effects because they measure 
three-particle correlations
\be
\la  \cos(\phi_a +\phi_b -2 \phi_{c}) \ra
= v_{1,a} v_{1,b} v_{2,c},
\label{cos3part}
\ee
in which there are no two-particle correlation terms and thus no non-flow 
contributions from them. The remaining non-flow is expected to cause a 
relative error of $20\%$, which is 
the major systematic uncertainty in this analysis.

\begin{figure}[htb]
\vspace*{-.3cm}
                 \insertplot{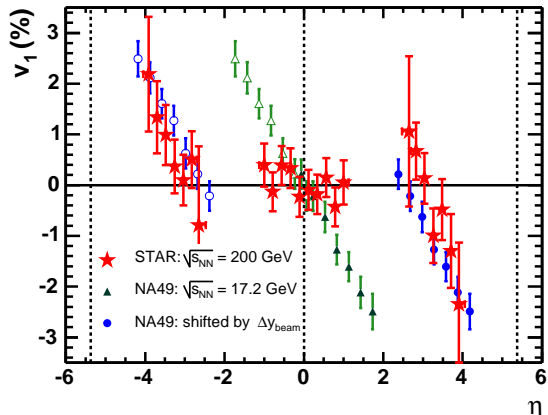}
\vspace*{-1.cm}
\caption[]{
The values of $v_1$ (stars) for charged 
particles for $10\%$ to $70\%$ centrality plotted as a function of 
pseudorapidity. Also shown are the results from NA49 (solid triangles) for 
pions from $158A$ GeV Pb+Pb midcentral ($12.5\%$ to $33.5\%$) collisions 
plotted as a function of rapidity. The measured NA49 points have also been 
shifted forward (solid circles) by the difference in the beam rapidities 
of the two accelerators. The open points have been reflected around 
midrapidity. The dashed lines indicate midrapidity and RHIC beam rapidity. 
Both results are from analyses involving three-particle cumulants, $v_1\{3\}$. This plot is taken from~\cite{v1v4}.
}
\label{v1}
\vspace*{-0.8cm}
\end{figure}

Fig.~\ref{v1} shows $v_1$ from three-particle cumulants ($v_1\{3\}$) 
along with corresponding results from NA49~\cite{NA49}. 
The RHIC $v_1(\eta)$ results differ greatly from the 
directly-plotted SPS data in that they are flat near midrapidity and only
become significantly different from zero at the highest rapidities measured. 
However, when the NA49 data is re-plotted in terms of rapidity relative to 
beam rapidity, they look similar. In the pseudorapidity region $|\eta|<1.2$,
$v_1(\eta)$ is approximately flat with a slope of $(-0.25 \pm 0.27)$\% per 
unit of pseudorapidity, which is consistent with predictions~\cite{wiggle}.  
Within errors we do not observe a wiggle in $v_1(\eta)$ at midrapidity. 
The quoted error is statistical only.
\vspace*{-0.1cm} 
\subsection{The evolution of elliptic flow}\label{ellipticFlow}
\vspace*{-0.1cm} 
It is interesting to see how elliptic flow evolves from p+p collisions, 
in which non-flow dominates, through d+Au collisions, where some 
correlation with the reaction plane might develop, and finally to Au+Au 
collisions, where flow dominates. To do such a comparison,
we calculate the azimuthal correlation of particles as a function 
of $p_t$ with the entire flow vector of all particles used to 
define the reaction plane (scalar product~\cite{Adler:2002pu}).  
The correlation in Au+Au collisions, under the assumption that 
non-flow effects in Au+Au collisions are similar to those in p+p 
collisions, is the sum of the flow and non-flow contribution and are 
given by:  
\be   
\la u_{t} Q^{*} \ra_{AA} = M_{AA} \, v_t \,  v_Q + \la u_{t} Q^{*} \ra_{pp},  
\label{eq:eQAA}  
\ee         
where $Q = \sum u_j$ and $Q^{*}$ its complex conjugate, $u_j 
=e^{2i\phi_j}$, $v_t$ is flow of particles with a given $p_t$, and 
$v_Q$ is the average flow of particles used to define $Q$.   
The first term in the r.h.s. of Eq.~\ref{eq:eQAA} represents the flow   
contribution; $M_{AA}$ is the multiplicity of particles contributing 
to the $Q$ vector.  This type of variable also can be extracted from the
cumulant approach~\cite{Borghini,OllitraultFlowWorkShop}: If we change the generating 
function that is used in the cumulant calculation~\cite{Borghini01} to
\be
G(z) = \prod_{j=1}^M
\left( 1+ {z^* u_j + z u_j^*} \right),      
\label{eq:newG0} 
\ee
where $z$ is an arbitrary complex number and $z^*$ denotes its complex 
conjugate. Then for a system that is a superposition of two independent 
system 1 and 2, and only ``non-flow'' correlations are present, we have 
\be
G(z) = G_1(z)G_2(z).
\label{eq:newG1} 
\ee
So if a nucleus-nucleus collision is a simple superposition of $N$ independent
p+p collisions, then
\be
G(z) = [ G_{pp}(z) ]^N.
\label{eq:newG2} 
\ee
We can readily see from Eq.(~\ref{eq:newG2}) that $Log(G(z))$ should 
scale linearly with $N$, so also should cumulants, which is the coefficient 
of $z$ of $Log(G(z))$. In the case of a second order cumulant, this is
\be
M^2 \la u_i u_j^* \ra = M \la u Q^* \ra,
\label{eq:newG3} 
\ee
dividing it by the scale factor (which is the multiplicity) one
recovers Eq. \ref{eq:eQAA} in the case if there is only non-flow.

The scalar product is a convenient quantity for this purpose because it 
is independent of 
multiplicity, which is very different in three collision systems. In the case 
of that only ``non-flow'' is present, scalar product should be the same 
for all three collision systems regardless of their system sizes. Any deviation
from fundamental p+p collisions for the scalar product results from collective
motions and/or effects from medium modification.

\begin{figure}
  \begin{center}
    \begin{minipage}[t]{0.48\linewidth}
\insertplot{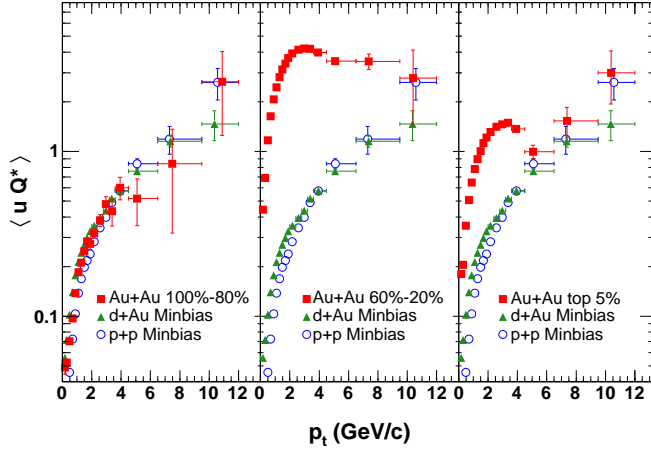}
    \end{minipage}\hfill
    \begin{minipage}[t]{0.29\linewidth}
    \vspace{0cm}  \caption{Azimuthal correlations in Au+Au collisions 
(squares) as a function of centrality (peripheral to central in the panels 
from left to right) compared to azimuthal correlations in minimum bias p+p 
collisions (circles) and d+Au collisions (triangles).  \label{fig:scalar}
}
    \end{minipage}
  \end{center}
\vspace{-0.5cm}
\end{figure}

Fig.~\ref{fig:scalar} shows the azimuthal correlation as a 
function of transverse momentum for  
three different centrality ranges in Au+Au collisions compared to 
minimum bias d+Au collisions and minimum bias p+p collisions.  
The difference at low $p_t$ between d+Au collisions 
and p+p collisions increases as a function of centrality (not shown) that 
is defined by the multiplicity in Au side, indicating that more collective motion 
is developed among soft particles in central d+Au collisions. 
This is consistent with Cronin effect in d+Au collisions, in which 
one expects that more scattering with soft particles is needed in order
to generate a relative high $p_t$ particle.
For Au+Au collisions, in middle central events we observe big deviation 
from p+p collisions that is due to the presence of elliptic flow,  
while in peripheral events, collisions are more like fundamental p+p 
collisions. At $p_t$ beyond $5$ GeV/c in central collisions, the azimuthal 
correlation in Au+Au collisions starts to follow that in p+p collisions, 
indicating a possible recovery of independent fragmentation. 
The centrality dependence of the azimuthal correlation 
in Au+Au collisions is clearly non-monotonic, being relatively small for very 
peripheral collisions, large for mid-central collisions, and relatively 
small again for central collisions. This non-monotonic centrality 
dependence is strong evidence that in mid-central collisions (60\%-20\%)
the measured finite $v_2$ for $p_t$ up to $7$ $\GeVc$ is due to 
real correlations with the reaction plane.

\vspace*{-0.1cm} 
\subsection{Higher harmonics}\label{higherHarmonics}
\begin{figure}
  \begin{center}
    \begin{minipage}[t]{0.48\linewidth}
\insertplot{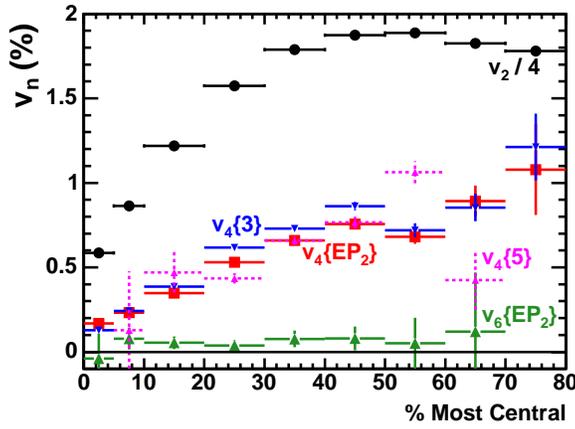}
    \end{minipage}\hfill
    \begin{minipage}[t]{0.38\linewidth}
    \vspace{0.5cm}  \caption{
The $p_t$- and $\eta$- integrated values of $v_2$,
$v_4$, and $v_6$ as a function of centrality. The $v_2$ values have
been divided by a factor of four to fit on scale. Also shown are the
three particle cumulant value ($v_4\{3\}$) and five particle cumulant value
($v_4\{5\}$). \label{fig:v4}
}
    \end{minipage}
  \end{center}
\vspace{-1cm}
\end{figure}

\vspace*{-0.1cm} 
Fig.~\ref{fig:v4} shows the centrality dependence for $p_t$-integrated
$v_2$, $v_4$, and $v_6$ with respect to the second harmonic event
plane and also $v_4$ from three-particle cumulants ($v_4\{3\}$).  The
five-particle cumulant, $v_4\{5\}$, is consistent with both methods but 
the error bars are about two times
larger. The $v_6$ values are close to zero for all centralities.
These results are averaged over $p_t$, thus reflecting mainly the low
$p_t$ region where the yield is large, and also averaged over $\eta$
for the midrapidity region accessible to the STAR TPC ($| \eta | \lt
1.2$). There has been a long history of searching for higher harmonics
and this is the first successful attempt of measuring higher harmonics
in heavy ion collisions. Such detailed measurement of the shape of the
event challenges in ever more detail the models describing the reaction.

\vspace*{-0.1cm} 
\section{Conclusions}\label{concl}
\vspace*{-0.1cm} 
  We have presented the first measurement of $v_1$
at RHIC energies. Within errors $v_1(\eta)$ is found to be approximately 
flat in the midrapidity region, which is consistent with microscopic transport
models, as well as hydrodynamical models where the flatness is
associated with the development of the expansion in the direction
opposite to the normal directed flow. Using the scalar product method, 
we studied the evolution of elliptic flow from elementary collisions 
(p+p) through collisions involving cold nuclear matter (d+Au), and then 
onto hot, heavy ion collisions (Au+Au). Measurements of higher harmonics 
are presented and discussed.

 
 

\vfill\eject
\end{document}